# Geometry and Topology in Relativistic Cosmology

Jean-Pierre Luminet

*Laboratoire Univers et Théories, CNRS-UMR 8102,*
*Observatoire de Paris, F-92195 Meudon cedex, France*

**Abstract.** General relativity does not allow one to specify the topology of space, leaving the possibility that space is multiply rather than simply connected. We review the main mathematical properties of multiply connected spaces, and the different tools to classify them and to analyse their properties. Following their mathematical classification, we describe the different possible muticonnected spaces which may be used to construct Friedmann-Lemaître universe models. Observational tests concern the distribution of images of discrete cosmic objects or more global effects, mainly those concerning the Cosmic Microwave Background. According to the 2003-2006 WMAP data releases, various deviations from the flat infinite universe model predictions hint at a possible non-trivial topology for the shape of space. In particular, a finite universe with the topology of the Poincaré dodecahedral spherical space fits remarkably well the data and is a good candidate for explaining both the local curvature of space and the large angle anomalies in the temperature power spectrum. Such a model of a small universe, whose volume would represent only about 80% the volume of the observable universe, offers an observational signature in the form of a predictable topological lens effect on one hand, and rises new issues on the physics of the early universe on the other hand.

## THE FOUR SCALES OF GEOMETRY

The forms which nature takes are limited by certain constraints. The first constraint is imposed by the three dimensional character of space (I am referring here to the usual three dimensions of length, width and depth, while being aware that recent theories invoke the existence of extra spatial dimensions which are only detectable on very small distance scales). Space is not a passive background, rather it has a structure which influences the shape of all existing objects. Every material form pays tribute to the rules dictated by the architecture of space.

The true architecture of space, and the constraints which it imposes, are still unknown. We can however reach a better understanding of the Universe by delving into the large range of abstract spaces arising in geometry, and by studying their local as well as their global structure. It is true that a mental image of non-Euclidean space eludes most of laymen, but geometry provides us with a consistent mathematical description.

Which mathematical space is capable of representing real physical space? The problem is much more complicated than it would appear. The microscopic and macroscopic worlds are profoundly different from the space of our immediate surroundings. The question of a geometric representation of space arises on four different levels, or, as the physicists say, four scales. These are microscopic, local, macroscopic and global.

On a local scale, that is to say for distances of between $10^{-18}$ metre (the distance now accessible to experimentation in particle accelerators) and $10^{11}$ metres (approximately the earth-sun distance), the geometry of space is very well described by that of ordinary, three dimensional Euclidean space $E^3$. « Very well » means that this mathematical structure serves as a natural framework for those physical theories, like Classical Mechanics and Special Relativity, which account satisfactorily for the quasi totality of natural phenomena.

On a macroscopic scale, that is to say for distances beween $10^{11}$ and $10^{25}$ metres, the geometry of space is better described as non-Euclidean, or, more accurately, as a continuous Riemannian manifold (a three dimensional generalisation of a surface with variable curvature). Such a space is curved to a greater or lesser extent by massive bodies (in the vicinity of exceptionally massive or dense bodies, like black holes, the effects of curvature can be felt over distances of a few metres only.) The physical framework is Einstein's General Theory of Relativity, in which the spacetime structure is more satisfactorily explained in terms of a supple, elastic fabric, gravitational phenomena being the manifestation of the non-zero curvature of the manifold.

On infinitesimally small distance scales, that is for distances less than $10^{-18}$ metre, we are into the realm of unexplored microscopic space. Neither powerful electron microscopes nor high energy particle accelerators can probe its most detailed structure. Here, geometric models only exist in the form of speculative theories. This microscopic space could reveal special geometric properties. What is it really made of ? Do « grains » of space, analogous to the grains of energy in Quantum Physics, actually exist ? Imaginative theorists, like Paul Dirac and John Wheeler, took this idea further by treating space like a collection of grains or soap bubbles. In their view, space is not simply a passive coordinate system. As a magical substance, whose curvature, granularity and excitations determine the masses, charges and fields of particles, it plays an active role in creating the material world. For example, space may be perturbed by fluctuations which permanently modify its shape, and make it extremely complicated - unstable, discontinuous and chaotic. It might even possess extra hidden dimensions.

These highly speculative topics, whose study is now well underway, will be among the stakes in tomorrow's Physics. I shall not be looking into them here, as we are mainly concerned with the widest perspective on the universal fabric of space. There, no lesser surprises await us. It is not yet known whether space is infinite, with zero or negative curvature; or whether it is finite, with a positive curvature, like a multidimensional sphere. Strangest of all would be a « wraparound » space, that is one folded back on itself. Such a space could be finite while being flat or negatively curved. Treating these global aspects of space requires a new discipline, a mixture of advanced mathematics and subtle cosmological observations : Cosmic Topology.

## CURVATURE VS. TOPOLOGY

The origins of topology go back to a riddle posed by the idle rich Prussians of the city of Königsberg, constructed around the branches of the Pregel river. The riddle consisted of deciding if, from any point in the city, it was possible to take a stroll in a closed loop while crossing once, and once only, each of the seven bridges which span the branches of the Pregel. The riddle was solved by the famous mathematician Leonhard Euler, who, in 1736, gave the necessary conditions which would allow such a route and, since the configuration of the bridges did not satisfy these rules, he proved that it was impossible to cross all seven bridges in a single trip.

The most important, Euler pointed out that, for the first time in the history of mathematics, one was dealing with a geometrical problem which had nothing to do with the metrics. The only important factors were the relative positions of the bridges. Indeed, if we trace the map of the city on a rubber sheet, and if we stretch or squeeze it in any direction without puncturing, cutting or tearing it, the nature of the problem is absolutely unchanged.

The solution given by Euler perfectly illustrates the two complementary aspects of geometry as the science of space: the « metric » part deals with the properties of distance, while the « topological » part studies the global properties, without introducing any measurements. The topological properties are those which remain insensitive to deformations, provided that these are continuous: with the condition of not cutting, piercing or gluing space, one can stretch it, crush it, or knead it in any way, and one will not change its topology, for example the fact that it is finite or infinite, the fact that it has holes or not, the number of holes if it has them, etc. It is easy to see that although continuous deformations may move the holes in a surface, they can neither create nor destroy them. Thus for a topologist, there is no difference between a rugby ball and a soccer ball. Worse, a ring and a coffee cup are one and the same object, characterized by a hole through which one can pass one's finger. On the other hand, a mug and a bowl are radically different on the level of topology, since a bowl does not have a handle.

Topology holds quite a few surprises. Let us take the Euclidean plane: it is an infinite two-dimensional page, that one visualizes most often within a three-dimensional space, although it has no need for this embedding to be perfectly well defined in an intrinsic way. The local geometry of the plane is determined by its metric, that is to say by the way in which lengths are measured. Here, it is sufficient to apply the Pythagorean theorem for a system of two rectilinear coordinates covering the plane : $ds^2 = dx^2 + dy^2$. This is a local measurement which says nothing about the finite or infinite character of space. Now let us change the topology. To do so, we take the plane and cut a strip of infinite length in one direction and finite width in the other. We then glue the two sides of the strip: we obtain a cylinder, a tube of infinite length. In this operation, the metric has not changed : the Pythagorean theorem still holds for the surface of the cylinder. The « intrinsic » curvature of the cylinder is therefore zero. This may appear surprising, since one has the impression that there is a non-zero curvature « somewhere », whose radius would be the radius of the cylinder. However, this « somewhere » calls into play a space exterior to the cylinder: the one in which we visualize it. In this sense, the cylinder has a so-called « extrinsic » curvature. Nevertheless, a flat creature, some sort of geometric paramecium living on the surface, would have access neither to this exterior space of higher dimension, nor to the extrinsic curvature of the cylinder. Tied to its two-dimensional space, it could make all of the necessary verifications (for instance measuring the sum of the angles in a triangle or the ratio of the circumference of a circle to its radius), and it would detect no difference with respect to the Euclidean metric of the infinite plane. The cylinder is said to be *locally Euclidean*.

Nevertheless, the cylinder differs from the plane in many respects. Certainly, its area is infinite, just like the plane, but it possesses a finite circumference in the direction perpendicular to its symmetry axis. In other words, the cylinder is anisotropic: not all directions are equivalent; following the length of a straight line parallel to the axis, one moves off toward infinity, while if one moves in the perpendicular direction, one returns to the departure point. In the operation of constructing a cylinder from a section of the plane, some of the global properties have changed; the cylinder thus has a different topology than that of the plane, while having the same metric. Its most remarkable characteristic is the existence of an infinite number of « straight lines » which join two arbitrary distinct points on the cylinder: those which make 0, 1, 2 . . . turns around the cylinder. Viewed in three dimensions, these straight lines are helices with constant spacing.

Let us continue with our cutting and gluing game. Take a tube of stretchable rubber, of finite length, and glue its two ends edge-to-edge. This is strictly equivalent to starting with a rectangle and gluing its opposite edges two by two. We obtain a torus, a surface having the shape of a ring or an inner tube. Here, a new difficulty arises. A real inner tube, just like the cylindrical tube, can be materialized in normal three-dimensional space; it therefore has an extrinsic curvature. However, in contrast to the tube, the inner tube also has a non-zero intrinsic curvature, which varies in different regions: sometimes positive, sometimes negative. However the toric surface obtained by identifying the opposite sides of a rectangle has an intrinsic curvature which is everywhere zero. This *flat torus*, a surface whose global properties are identical to those of a ring but whose curvature is everywhere zero, cannot be viewed within our usual three-dimensional space (it can only be embedded into $E^4$). Yet one can describe all of its properties without exception; its area is finite in the sense that it is impossible to move infinitely far away from one's departure point, and it is not isotropic, since two of its directions, named the principal directions, are privileged. Let us imagine a creature living on a flat torus, moving straight ahead along a principal direction; she communicates via light rays with her departure point, in such a way that she can calculate the distance traveled; at a certain moment, this distance attains a maximum, and then begins to decrease; after having made a complete circuit, the creature has returned to her point of departure. She would conclude from this that she lives in a space of finite extent. Nevertheless, by having measured the sum of the angles in a triangle in these surroundings, she has still found 180 degrees, because of which she would also deduce that she lives in a Euclidean plane. The metric (local geometry) of the flat torus is still given by the Pythagorean theorem, just like that of the plane and the cylinder.

Through simple cutting and re-gluing of parts of the plane, we have thus defined two surfaces with different topologies than the plane : the cylinder and the flat torus, which however belong to the same family, the locally Euclidean surfaces. The gluing method becomes extremely fruitful when the surfaces are more complicated. Let us take two tori and glue them to form a « double torus ». As far as its topological properties are concerned, this new surface with two holes can be represented as an eight-sided polygon (an octagon), which can be understood intuitively by the fact that each torus was represented by a quadrilateral. But this surface is not capable of paving the Euclidean plane, for an obvious reason: if one tries to add a flat octagon to each of its edges, the eight octagons will overlap each other. One must therefore curve in the sides and narrow the angles, in other words pass to a hyperbolic space: only there does one succeed in fitting eight octagons around the central octagon, and starting from each of the new octagons one can construct eight others, *ad infinitum*. By this process one paves an infinite space : the Lobachevsky hyperbolic plane.

More generally, a two-dimensional n-torus $T_n$ is a torus with *n* holes. $T_n$ can be constructed as the connected sum of *n* simple tori. The n-torus is therefore topologically equivalent to a connected sum of *n* squares whose opposite edges have been identified. This sum is itself topologically equivalent to a 4n-gon where all the vertices are identical with each other and the sides are suitably identified by pairs. Such an operation is not straightforward when *n ≥ 2*. All the vertices of the polygon correspond to the same point of the surface. Since the polygon has at least 8 edges, it is necessary to make the internal angles thinner in order to fit them suitably around a single vertex. This can only be achieved if the polygon is represented in the hyperbolic plane $H^2$ instead of the Euclidean plane $E^2$ : this increases the area and decreases the angles. The more angles to fit together, the thinner they have to be and the greater the surface. The n-torus (*n ≥ 2*) is therefore a compact surface of negative curvature. This type of surface is most commonly seen at bakeries, in the form of pretzels. We call them « hyperbolic pretzels ». They all have the same local geometry, of hyperbolic type ; however, they do not have the same topology, which depends on the number of holes.

When one deals with more than two dimensions, the gluing method remains the simplest way to visualize spaces. By analogy with the two-dimensional case, the three-dimensional simple torus $T^3$ (also referred to as the *hypertorus*) is obtained by identifying the opposite faces of a parallelepiped. The resulting volume is finite. Let us imagine a light source at our position, immersed in such a structure. Light emitted backwards crosses the face of the parallelepiped behind us and reappears on the opposite face in front of us; therefore, looking forward we can see our back. Similarly, we see in our right our left profile, or upwards the bottom of our feet. In fact, for light emitted isotropically, and for an arbitrarily large time to wait, we could observe ghost images of any object viewed arbitrarily close to any angle. The resulting visual effect would be comparable (although not identical) to what could be seen from inside a parallelepiped of which the internal faces are covered with mirrors. Thus one would have the visual impression of infinite space, although the real space is closed.

# BASICS OF TOPOLOGY

**Simple vs. multiple connectedness**

Let us now formalize a little bit more the topological notions introduced above. The strategy for characterizing the shape of a space *M* is to produce invariants which capture the key features of the topology and uniquely specify each equivalence class. The topological invariants can take many forms. They can be just numbers, such as the dimension of the manifold, the degree of connectedness or the Poincaré-Euler characteristic. They can also be whole mathematical structures, such as the homotopy groups. The latter are defined in an elegant way from the tightening of laces. A lace is a closed curve traced on a surface. On the infinite plane, we can draw an arbitrary lace, however large, from an arbitrary point; this lace can always be retightened and reduced to a point without encountering any obstacles. The topologists call such a space *simply connected*. Formally, a lace at *x* in *M* is any path which starts at *x* and ends at *x*. Two laces *g* and *g'* are homotopic if *g* can be continuously deformed into *g'*. The manifold *M* is simply-connected if every lace is homotopic to a point. Obviously, the Euclidean spaces $E^1$, $E^2$, ... $E^n$, and the spheres $S^2$, $S^3$, ... $S^n$ are simply-connected.

On the other hand, the circle $S^1$, the cylinder $S^1 x E^1$ and the torus $S^1 x S^1$ do not have this property. Of course, there are laces which can be completely retightened, as in the plane; but some of them cannot: a circle which wraps around the cylinder or which is traced around the torus, for example, cannot be continuously shrunk to a point. For such spaces, the topology is said to be *multiply connected*.

The study of homotopic laces in a manifold *M* is a way of detecting holes or handles. Moreover the equivalence classes of homotopic laces can be endowed with a group structure, essentially because laces can be added by joining them end to end. The group of laces is called the first homotopy group at *x* or, in the terminology originally introduced by Poincaré, the fundamental group $\pi_1(M, x)$. The fundamental group is independent of the base point: it is a topological invariant of the manifold.

For surfaces, multi-connectedness means that the fundamental group is non trivial: there is at least one lace that cannot be shrunk to a point. But in higher dimensions the problem is more complex because laces, being only one-dimensional structures, are not sufficient to capture all the topological features of the manifolds. The purpose of algebraic topology, extensively developed during the twentieth century, is to generalise the concept of homotopic laces and to define higher homotopy groups. However the fundamental group (the first homotopy group) remains essential.

**Fundamental domain and holonomy group**

In the nineteenth century, mathematicians discovered that it is possible to represent any surface whatsoever with a polygon whose sides one identifies, two by two. The torus is topologically equivalent to a rectangle with opposite edges identified. The rectangle is called a *fundamental domain* (hereafter FD) of the torus. From a topological point of view (namely without reference to size), the FD can be chosen in different ways: a square, a rectangle, a parallelogram, even a hexagon (since the plane can be tiled by hexagons, the flat torus can be also represented by a hexagon with suitable identification of edges).

The FD distinctly characterizes a certain aspect of the topology. But this is not enough; we must also specify the geometric transformations which identify the points. Indeed, starting from a square, one could identify the points diametrically opposite with respect to the center of symmetry of the square, and the surface obtained will no longer be a flat torus; it will no longer even be Euclidean, but spherical, a surface called the projective plane. The mathematical transformations used to identify points form a group of symmetries, called the *holonomy group*.

This group is discrete, i.e. there is a non zero shortest distance between any two homologous points, and the generators of the group (except the identity) have no fixed point. This last property is very restrictive (it excludes for instance the rotations) and allows the classification of all possible holonomy groups. Due to the fact that the holonomy group is discrete, the FD is always convex and has a finite number of faces. In two dimensions, it is a surface whose boundary is constituted by lines, thus a polygon. In three dimensions, it is a volume bounded by faces, thus a polyhedron.

**Universal covering**

Starting from the fundamental domain and acting with the transformations of the holonomy group on each point, one creates a number of replicas of the FD ; we produce a sort of tiling of a larger space, called the *universal covering space* (hereafter UC) ***M*****. By construction, ***M*** is locally indistinguishable from ***M***. But its topological properties can be quite different. The UC is necessarily simply connected : any lace can be shrunk to a point. Thus, when ***M*** is simply-connected, it is identical to its universal covering space ***M*****. But when ***M*** is multiply connected, each point of ***M*** generates replicas of points in ***M*****. The universal covering space can be thought of as an unwrapping of the original manifold. For instance, the UC of the flat torus is the Euclidean plane ***E**$^2$*, which indeed reflects the fact that the flat torus is a locally Euclidean surface.

**Spaceforms**

To summarize : the shape of a homogeneous space is entirely specified if one is given a fundamental domain; a particular group of symmetries, the holonomies, which identify the edges of the domain two by two; and a universal covering space that is paved by fundamental domains. Classifying the possible shapes thus reduces, in part, to classifying symmetries.

Let us apply this recipe in order to list all homogeneous surfaces: two-dimensional spaces with no boundaries and no sharp points. As far as the curvature is concerned, homogenous surfaces are of three types: spherical surfaces, with positive curvature (like the surface of a rugby ball); Euclidean surfaces, with zero curvature (whose planar geometry is taught in high school); and hyperbolic surfaces, of negative curvature (like certain parts of a saddle or of a trumpet's horn). Within each of these basic types, mathematicians have classified all possible topologies – also referred to as *spaceforms*.

There are only two forms for spherical surfaces, both of them finite: the sphere, which can be given a wide variety of different metrical aspects depending on what sort of continuous stretching it is subjected to, and the projective plane. The sphere is simply connected, the projective plane is not. The latter surface is not easily visualized; the simplest way to do so is to pass through the intermediary of its fundamental domain, a disk, whose diametrically opposite points are identified.

Euclidean surfaces can come in five possible shapes: the plane, of course, which is the simply connected prototype, but also the cylinder, the Möbius band (which is an infinitely wide Möbius strip), the flat torus, and the Klein bottle, all of which are multiply connected. The first three are infinite, the other two finite. These surfaces, although conceptually simple, are not all easy to visualize; thus, although the Klein bottle has no curvature, it is closed in on itself and has neither inside nor outside; it is said to be « non-orientable ».

Finally, the hyperbolic surfaces, with negative curvature, have an infinite number of topologies. Only one of them, equivalent to the Lobachevsky plane, is simply connected. All others are multiply connected, characterized by the number of holes. We have seen, for example, that the surface of a generalized pretzel is hyperbolic.

One conclusion that we can quickly draw from this classification is that, in the infinite set of homogeneous surfaces, they are all hyperbolic, up to only seven exceptions.

# THREE-DIMENSIONAL MANIFOLDS OF CONSTANT CURVATURE

The passage from two dimensions to three dimensions in no way reduces to a simple generalization, but leads to the appearance of radically new properties. Every regular surface can be homogenized so as to be described by a metric of constant curvature ; this means that there are only three prototypical simply connected surfaces (which serve as universal coverings), to which all other surfaces are necessarily related. Things are not the same for three-dimensional spaces : there are eight possible universal covering spaces (see Thurston, 1997, for a synthesis). Only three of these are homogeneous and isotropic, the remaining five are homogeneous but not isotropic, meaning that at a given point the measurement of the curvature depends on direction.

Three-dimensional cylinders are some relatively simple examples of these. In the same way that the usual cylinder can be considered as the « product » $S^1 \times E^1$ of a circle $S^1$ and a straight line $E^1$ (in the sense that if one slides a circle along a straight line perpendicular to its center one creates a cylinder), the « three-dimensional spherical cylinder » can be pictured as the product $S^2 \times E^1$ of a sphere $S^2$ and a straight line $E^1$. However, while the cylindrical surface could be described with the metric of the Euclidean plane $E^2$, the cylindrical-spherical space is fundamentally distinct from the Euclidean space $E^3$. The curvatures measured are different depending on the orientations of the referential planes used to cut it. Similarly, the « cylindrical-hyperbolic » space $H^2 \times E^1$, obtained by stacking Lobachevsky planes, is fundamentally distinct from $E^3$.

Cosmology, however, focuses mainly on locally homogeneous and isotropic spaces, namely those admitting one of the three geometries of constant curvature. Any compact 3-manifold $M$ with constant curvature $k$ can thus be expressed as the quotient $M = M^*/\Gamma$, where the universal covering space $M^*$ is either :
- the Euclidean space $E^3$ if $k = 0$
- the 3-sphere $S^3$ if $k > 0$
- the hyperbolic 3-space $H^3$ if $k < 0$

and $\Gamma$ is a subgroup of isometries of $M^*$ acting freely and discontinuously.

**Euclidean space forms**

Simply-connected Euclidean space, $E^3$, with uniformly zero curvature, is infinite in every direction. Its full isometry group is $G = ISO(3) = E^3 \times SO(3)$, and the generators of the possible holonomy groups $\Gamma$ (i.e., discrete subgroups without fixed point) include the identity, the translations, the glide reflections and the screw motions (combinations of a rotation and a translation parallel to the axis of rotation) occurring in various combinations. The multiply connected Euclidean spaces are characterized by their fundamental polyhedra and their holonomy groups. The fundamental polyhedra are either a finite or infinite parallelepiped, or a prism with a hexagonal base, corresponding to the two ways of tiling Euclidean space. The various different combinations generate seventeen distinct multiply connected Euclidean spaces (for an exhaustive study, see Riazuelo et al., 2004a).

Seven of these spaces are open (of infinite volume). Two of these, called *slab spaces*, are made of a slab that extends infinitely in two directions, but has finite thickness. The two ends are identified by a translation or after a rotation of 180°. The five others, called *chimney spaces*, are made of a rectangular chimney of infinite height, whose front and back (and left and right) surfaces are identified by a translation and appropriate rotations.

Ten other Euclidean spaces are closed (of finite volume). The first six spaces are orientable hypertori. The simplest hypertorus $T^3$ is constructed by identifying the opposite faces of a parallelepiped by translations. The other hypertori are obtained after gluing with a quarter turn, a half-turn, a one-sixth turn and a one-third turn, while the Hantzsche-Wendt space has a more complicated structure. It is these six compact, orientable Euclidean spaces that present a particular interest for cosmology, since they could perfectly model the spatial part of the so-called « flat » universe models.

Eventually, four closed Euclidean spaces are non-orientable generalizations of the Klein bottle : Klein space, Klein space with a horizontal flip, Klein space with a vertical flip and Klein space with a half-turn.

**Spherical space forms**

The simply-connected spherical space $S^3$, with positive curvature, is the hypersphere. Einstein attempted to give an intuitive image of such a finite yet limitless three-dimensional space, that a little bit of exercise suffices to render familiar to our thinking. A way to visualize the hypersphere consists in imagining the points of the hypersphere as those of a family of two-dimensional spheres which grow in radius from 0 to a maximal value $R$, then shrink from $R$ back to 0 (in the same way that a sphere can be cut into planar slices which are circles of varying radius). Another possibility is to view the hypersphere as composed of two spherical balls embedded in Euclidean space, glued along their boundaries in such a way that each point on the boundary of one ball is the same as the corresponding point on the other ball.

The full isometry group of $S^3$ is SO(4). The holonomies that preserve the metric of the hypersphere, i.e. the admissible subgroups G of SO(4) without fixed point, acting freely and discontinuosly on $S^3$, belong to three categories :

- the cyclic groups of order $p$, $Z_p$ $(p \geq 2)$, made up of rotations by an angle $2\pi/p$ around a given axis, where $p$ is an arbitrary integer ;
- the dihedral groups of order $2m$, $D_m$ $(m>2)$, which are the symmetry groups of a regular plane polygons of $m$ sides;
- the binary polyhedral groups, which preserve the shapes of the regular polyhedra. The group $T^*$ preserves the tetrahedron (4 vertices, 6 edges, 4 faces), of order 24 ; the group $O^*$ preserves the octahedron (6 vertices, 12 edges, 8 faces), of order 48 ; the group $I^*$ preserves the icosahedron (12 vertices, 30 edges, 20 faces), of order 120. There are only three distinct polyhedral groups for the five polyhedra, because the cube and the octahedron on the one hand, the icosahedron and the dodecahedron on the other hand are duals, so that their symmetry groups are the same.

If one identifies the points of the hypersphere by holonomies belonging to one of these groups, the resulting space is spherical and multiply connected. For an exhaustive classification, see Gausmann et al. (2001). There is a countable infinity of these, because of the integers $p$ and $m$ which parametrize the cyclic and dihedral groups.

The spaces with cyclic group are called *lens spaces*, denoted thus because their fundamental polyhedra have the shapes of lenses. For instance, the projective (also called elliptic) space $P^3 = S^3/Z_2$ is obtained by identifying diametrically opposite points on $S^3$. It was used by de Sitter (1917) and Lemaître (1931) as the space structure of their cosmological models, while Einstein (1917) selected the symply connected hypersphere.

The spaces with dihedral group are called *prism spaces*, because of the shape of their fundamental polyhedra. Finally, the spaces with polyhedral groups are called *polyhedral spaces*. Among them, the *Poincaré Dodecahedral Space $S^3/I^*$* is obtained by identifying the opposite pentagonal faces of a regular spherical dodecahedron after rotating by 1/10[th] turn in the clockwise direction around the axis orthogonal to the face. This configuration involves 120 successive operations and gives some idea of the extreme complication of such multiply connected topologies. Its volume is 120 times smaller than that of the hypersphere with the same radius of curvature, and it is of particular interest for cosmology, giving rise to fascinating topological mirages (see below).

Since the universal covering $S^3$ is compact, all the multiply connected spherical spaces are also compact. As the volume of $S^3$ is $2\pi^2 R^3$, the volume of $M = S^3/\Gamma$ is simply $vol(M) = 2\pi^2 R^3 /|\Gamma|$ where $|\Gamma|$ is the order of the group $\Gamma$. For topologically complicated spherical 3-manifolds, $|\Gamma|$ becomes large and $vol(M)$ is small. There is no lower bound since $\Gamma$ can have an arbitrarily large number of elements (for lens and prism spaces, the larger $p$ and $m$ are, the smaller the volume of the corresponding spaces). Hence $0 < vol(M) \leq 2\pi^2 R^3$. In contrast, the diameter, i.e., the maximum distance between two points in the space, is bounded below by $\sim 0.326\,R$, corresponding to the dodecahedral space.

**Hyperbolic space forms**

Locally hyperbolic manifolds are less well understood than the other homogeneous spaces. However, according to the pioneering work of Thurston, almost all 3-manifolds can be endowed with a hyperbolic structure. The universal covering space, $H^3$, is the three-dimensional analog of the Lobachevsky plane $H^2$, and extends to infinity in every direction. Its group of isometries is isomorphic to PSL(2,C), namely the group of fractional linear transformations acting on the complex plane. Finite subgroups are discussed in Beardon (1983). The mathematicians have not succeeded in classifying all of them, but they know an infinite number of examples. Some of these spaces are closed (with finite volume), and others are open (with infinite volume).

In hyperbolic geometry there is an essential difference between the 2-dimensional case and higher dimensions. A surface of genus $g \geq 2$ supports uncountably many non equivalent hyperbolic metrics. But the so-called *rigidity theorem* proves that a connected oriented n-dimensional manifold supports *at most one* hyperbolic metric as soon as $n \geq 3$. In simple terms, this means that if one fixes a hyperbolic topology, there is only a single metric compatible with this topology. From this, it follows that the volume of space (in units of the curvature radius $R$) is fixed by its topology. It is thus possible to classify the closed hyperbolic spaces by increasing volumes, which could have seemed, at a first glance, contradictory with the very purpose of topology.

However the volumes cannot be made arbitrarily small by gluing operations. The absolute lower bound is $V_{min} = 0.16668$, but no space has been constructed having precisely this volume. Until now, the smallest known hyperbolic space (that is to say one whose fundamental polyhedron and holonomy group were able to be completely calculated) is *Weeks space*, with a volume equal to 0.94272. Its FD is a polyhedron with 26 vertices and 18 faces, of which 12 are pentagons and 6 are quadrilaterals. Its outer structure, the Klein coordinates of the vertices and the 18 matrix representations of the generators of the holonomy group are given in Lehoucq et al. (1999).

In cosmology, the Weeks manifold leaves room for many topological lens effects, since the volume of the observable universe is about 200 times larger than the volume of Weeks space for $\Omega_0 = 0.3$. Indeed, any compact hyperbolic space have geodesics shorter than the curvature radius, leaving room to fit a great many copies of a fundamental polyhedron within the horizon radius, even for manifolds of volume ~10. The publicly available program SnapPea, available on the internet (Weeks) classifies all known spaces by increasing volumes, and gives their properties: the structure of the fundamental polyhedron, the nature of the transformations in the holonomy group, the characteristic topological lengths, etc. Several millions of compact hyperbolic spaces with volume less than 10 could be calculated.

# TOPOLOGY AND COSMOLOGY

General relativity has successfully passed a number of experimental tests, but, like any physical theory, it is incomplete. One of the limits on its validity is well known: it does not take into account the microscopic properties of matter, described by quantum physics. Einstein was well aware of this, since, after putting the finishing touches on his gravitational theory in 1916, he passed the rest of his days attempting to unify gravity with the other physical interactions, in vain. Present day attempts at unification, whether « superstrings », « M-theory » or « quantum loop gravity », tend to run into the same difficulties (see e.g. Smolin, 2002). What is less known is that general relativity is also incomplete on the large scale: is space finite or infinite, oriented or not? What is its global shape? Gravitation does not by itself decide the overall form taken by space. The preceding examples have indeed shown that the curvature of space does not necessarily allow one to come to any conclusions about its finite or infinite character.

These basic cosmological questions come from the global topology of the Universe, about which general relativity is silent. Einstein's theory in fact only allows one to deal with the local geometric properties of the Universe. Its partial differential equations have as a solution a metric tensor $g_{ab}$, or, equivalently, the infinitesimal element of distance $ds^2$ separating two events in space-time. This leads the study of the Universe, of its content, and of its physical properties to problems of differential geometry on a pseudo-Riemannian manifold.

It is presently believed that our Universe is correctly described at large scale by a Friedmann-Lemaître (hereafter FL) model. The FL models are homogeneous and isotropic solutions of Einstein's equations, of which the spatial sections have constant curvature ; they include the de Sitter solution, as well as those incorporating a cosmological constant, or a non standard equation of state. The FL models fall into 3 general classes, according to the sign of their spatial curvature $k = -1, 0,$ or $+1$. The spacetime manifold is described by the metric $ds^2 = c^2 dt^2 - R^2(t) d\sigma^2$, where $d\sigma^2 = d\chi^2 + S_k^2(\chi) (d\theta^2 + \sin^2\theta\, d\phi^2)$ is the metric of a 3-dimensional homogeneous manifold, flat [$k=0$] or with curvature [$k \pm 1$]. The function $S_k(\chi)$ is defined as $sinh(\chi)$ if $k= -1$, $\chi$ if $k=0$, $sin(\chi)$ if $k=1$; $R(t)$ is the scale factor, chosen equal to the spatial curvature radius for non flat models.

The spatial topology is usually assumed to be the same as that of the corresponding simply connected, universal covering space: the hypersphere, Euclidean space or the 3D-hyperboloid, the first being finite and the other two infinite. However, there is no particular reason for space to have a simply connected topology. In any case, general relativity says nothing on this subject; it is only the strict application of the cosmological principle, added to the theory, which encourages a generalization of locally observed properties to the totality of the Universe. Likewise, an ant in the middle of the desert would be convinced that the entire world is made of grains of sand. However, to the metric element given above there are several, if not an infinite number, of possible topologies, and thus of possible models for the physical Universe. For example, the hypertorus and familiar Euclidean space are locally identical, and relativistic cosmological models describe them with the same FL equations, even though the former is finite and the latter infinite ; likewise, the equations for a Universe of negative curvature make no distinction between a finite or an infinite space. In fact, only the boundary conditions on the spatial coordinates are changed. Thus the multi-connected cosmological models share exactly the same kinematics and dynamics as the corresponding simply connected ones (for instance, the time evolution of the scale factor $R(t)$ is identical).

At this stage, it is wise to recall that cosmological models do not reduce to three dimensions, but are four dimensional space-times. Thus, to the problem of the topology of space is added that of the topology of time. What can be said about the space-time topology ? An infinite spectrum of possibilities offer themselves as models. Nevertheless, some brief consideration of the physical properties of the Universe allows us to rapidly isolate a good number of inadmissible topologies. Here is why. Models of the big bang are homogeneous, meaning that their spatial part has a curvature which is everywhere uniform, and expanding. These two properties allow one to unambiguously distinguish slices of simultaneous space and the axis of cosmic time. We can therefore describe space-time as the mathematical product of a three-dimensional Riemannian space and the time axis. This foliation considerably simplifies things. Time is represented by a 1D-space whose points represent instants: a single number suffices to determine a particular time. Time possesses an ordered structure: on a line, one point is necessarily situated either before or after another point. The topology of time is in the end rather poor; in contrast to that of multidimensional space, it only offers two cases : the line $E^1$ and the circle $S^1$. These two forms in fact correspond to two great philosophical conceptions, linear time and cyclic time. The latter has long prevailed in myths, such as that of the Eternal Return, but today it has been abandoned by physics because it violates the principle of causality, according to which cause must precede effect. As a consequence, any identification of points along the time axis is forbidden. In the framework of cosmological models of expansion followed by contraction, one could, certainly, think to identify the big bang and the big crunch, that is to say the beginning of time with its end; but this operation is unlawful, for these points are singularities which are not even part of the Universe.

The question of cosmic topology therefore reduces principally to that of the spatial component of the Universe. For each type of possible curvature, as we have seen, there are various FL models with multiply connected topologies. In relativistic cosmology, the curvature of physical space depends on the way the total energy density of the Universe may counterbalance the kinetic energy of the expanding space. The normalized density parameter $\Omega_0$, defined as the ratio of the actual density to the critical value that an Euclidean space would require, characterizes the present-day contents (matter, radiation and all forms of energy) of the Universe. If $\Omega_0$ is greater than 1, then space curvature is positive and geometry is spherical; if $\Omega_0$ is smaller than 1 the curvature is negative and geometry is hyperbolic; eventually $\Omega_0$ is strictly equal to 1 and space is Euclidean.

The next question about the shape of the Universe is to know whether space is finite or infinite – equivalent to know whether space contains a finite or an infinite amount of matter–energy, since the usual assumption of homogeneity implies a uniform distribution of matter and energy through space. From a purely geometrical point of view, all positively curved spaces are finite whatever their topology, but the converse is not true : flat or negatively curved spaces can have finite or infinite volumes, depending on their degree of connectedness (Ellis, 1971 ; Lachièze-Rey & Luminet, 1995).

From an astronomical point of view, it is necessary to distinguish between the "observable universe", which is the interior of a sphere centered on the observer and whose radius is that of the cosmological horizon (roughly the radius of the last scattering surface), and the physical space. There are only three logical possiblities. First, the physical space is infinite – like for instance the simply-connected Euclidean space. In this case, the observable universe is an infinitesimal patch of the full universe and, although it has long been the preferred model of many cosmologists, this is not a testable hypothesis. Second, physical space is finite (e.g. an hypersphere or a closed multiconnected space), but greater than the observable space. In that case, one easily figures out that if physical space is much greater that the observable one, no signature of its finitude will show in the observable data. But if space is not too large, or if space is not globally homogeneous (as is permitted in many space models with multiconnected topology) and if the observer occupies a special position, some imprints of the space finitude could be observable. Third, physical space is smaller than the observable universe. Such an apparently odd possibility is due to the fact that space can be multiconnected and have a small volume. There is a lot of possibilites, whatever the curvature of space. Small universe models may generate multiple images of light sources, in such a way that the hypothesis can be tested by astronomical observations. The smaller the fundamental domain, the easier it is to observe the multiple topological imaging. Lehoucq et al. (1998) have calculated, for a given catalog of observable cosmic sources (discrete or diffuse) with a given depth in redshift, the approximate number of topological images in locally hyperbolic and locally spherical spaces as a function of the cosmological paramaters $\Omega_m$ and $\Omega_\Lambda$. How do the present observational data constrain the possible multi-connectedness of the universe and, more generally, what kinds of tests are conceivable ? The following sections deal with these matters (see Luminet, 2001 for a non-technical book about all the aspects of topology and its applications to cosmology).

## THE DRUMHEAD UNIVERSE

The topology and the curvature of space can be studied by using specific astronomical observations. For instance, from Einstein's field equations, the space curvature can be deduced from the experimental values of the total energy density and of the expansion rate. If the Universe was finite and small enough, we should be able to see

« all around » it, because the photons might have crossed it once or more times. In such a case, any observer might identify multiple images of a same light source, although distributed in different directions of the sky and at various redshifts, or to detect specific statistical properties in the apparent distribution of faraway sources such as galaxy clusters. To do this, methods of « cosmic crystallography » have been devised (Lehoucq et al., 1996 ; Uzan et al., 1999), and extensively studied by the Brazilian school of cosmic topology (see e.g. Gomero et al., 2002). The main limitation of cosmic crystallography is that the presently available catalogs of observed sources at high redshift are not complete enough to perform convincing tests.

Fortunately, the topology of a small Universe may also be detected through its effects on such a Rosetta stone of cosmology as is the cosmic microwave background (hereafter CMB) fossil radiation (Levin, 2002). If you sprinkle fine sand uniformly over a drumhead and then make it vibrate, the grains of sand will collect in characteristic spots and figures, called Chladni patterns. These patterns reveal much information about the size and the shape of the drum and the elasticity of its membrane. In particular, the distribution of spots depends not only on the way the drum vibrated initially but also on the global shape of the drum, because the waves will be reflected differently according to whether the edge of the drumhead is a circle, an ellipse, a square, or some other shape. In cosmology, the early Universe was crossed by real acoustic waves generated soon after the big bang. Such vibrations left their imprints 380 000 years later as tiny density fluctuations in the primordial plasma. Hot and cold spots in the present-day 2.7 K CMB radiation reveal those density fluctuations. Thus the CMB temperature fluctuations look like Chladni patterns resulting from a complicated three-dimensional drumhead that vibrated for 380 000 years. They yield a wealth of information about the physical conditions that prevailed in the early Universe, as well as present geometrical properties like space curvature and topology. More precisely, density fluctuations may be expressed as combinations of the vibrational modes of space, just as the vibration of a drumhead may be expressed as a combination of the drumhead's harmonics. The shape of space can be heard in a unique way. Lehoucq et al. (2002) calculated the harmonics (the so-called "eigenmodes of the Laplace operator") for most of the spherical topologies, and Riazuelo et al. (2004a) did the same for all 18 Euclidean spaces. Then, starting from a set of initial conditions fixing how the universe originally vibrated (the so-called Harrison-Zeldovich spectrum), it is possible to evolve the harmonics forward in time to simulate realistic CMB maps for a number of flat and spherical topologies (Uzan et al., 2004).

The "concordance model" of cosmology describes the Universe as a flat infinite space in eternal expansion, accelerated under the effect of a repulsive dark energy. The data collected by the NASA satellite WMAP (Spergel et al., 2003) have produced a high resolution map of the CMB which showed the seeds of galaxies and galaxy clusters and allowed to check the validity of the dynamic part of the expansion model. However, combined with other astronomical data (Tonry et al., 2003), they suggest a value of the density parameter $\Omega_0 = 1.02 \pm 0.02$ at the 1σ level. The result is marginally compatible with strictly flat space sections. Improved measurements could indeed lower the value of $\Omega_0$ closer to the critical value 1, or even below to the hyperbolic case. Presently however, taken at their face value, WMAP data favor a positively curved space, necessarily of finite volume since all spherical spaceforms possess this property.

Now what about space topology ? There is an intriguing feature in WMAP data, already present in previous COBE mearurements (Hinshaw et al., 1996), although at a level of precision that was not significant enough to draw firm conclusions. The power spectrum depicts the minute temperature differences on the last scattering surface, depending on the angle of view. It exhibits a set of peaks when anisotropy is measured on small and mean scales (i.e. concerning regions of the sky of relatively modest size). These peaks are remarkably consistent with the infinite flat space hypothesis. At large angular scales, the concordance model predicts that the power spectrum should follow the so-called "Sachs-Wolfe plateau". However, WMAP measurements fall well below the plateau for the quadrupole and the octopole moments (i.e. for CMB spots typically separated by more than 60°). Since the flat infinite space model cannot explain this feature, it is necessary to look for an alternative.

CMB temperature anisotropies essentially result from density fluctuations of the primordial Universe : a photon coming from a denser region will loose a fraction of its energy to compete against gravity, and will reach us cooler. On the contrary, photons emitted from less dense regions will be received hotter. The density fluctuations result from the superposition of acoustic waves which propagated in the primordial plasma. Riazuelo et al. (2004a) have developed complex theoretical models to reproduce the amplitude of such fluctuations, which can be considered as vibrations of the Universe itself. In particular, they simulated high resolution CMB maps for various space topologies (Riazuelo et al., 2004b) and were able to compare their results with real WMAP data. Depending on the underlying topology, the distribution of the fluctuations differs. For instance, in an infinite flat space, all wavelengths are allowed, and fluctuations must be present at all scales.

The CMB temperature fluctuations can be decomposed into a sum of *spherical harmonics*, much like the sound produced by a music instrument may be decomposed into ordinary harmonics. The "fundamental" fixes the height of the note (as for instance a 440 hertz acoustic frequency fixes the *A* of the pitch), whereas the relative amplitudes of each harmonics determine the tone quality (such as the *A* played by a piano differs from the *A* played by a harpsichord). Concerning the relic radiation, the relative amplitudes of each spherical harmonics determine the power spectrum, which is a signature of the space geometry and of the physical conditions which prevailed at the time of CMB emission.

The first observable harmonics is the quadrupole (whose wavenumer is $l = 2$). WMAP has observed a value of the quadrupole 7 times weaker than expected in a flat infinite Universe. The probability that such a discrepancy occurs by chance has been estimated to 0.2 % only. The octopole (whose wavenumber is $l = 3$) is also weaker (72 % of the expected value). For larger wavenumbers up to $l = 900$ (which correspond to temperature fluctuations at small angular scales), observations are remarkably consistent with the standard cosmological model.

The unusually low quadrupole value means that long wavelengths are missing. Some cosmologists have proposed to explain the anomaly by still unknown physical laws of the early universe (Tsujikawa et al., 2003). A more natural explanation may be because space is not big enough to sustain long wavelengths. Such a situation may be compared to a vibrating string fixed at its two extremities, for which the maximum wavelength of an oscillation is twice the string length. On the contrary, in an infinite flat space, all the wavelengths are allowed, and fluctuations must be present at all scales. Thus this geometrical explanation relies on a model of finite space whose size *smaller* than the observable universe constrains the observable wavelengths below a maximum value.

Such a property has been known for a long time, and was used to constrain the topology from COBE observations (Sokolov, 1993). Preliminary oversimplified analyses (de Oliveira-Costa & Smoot, 1995) suggested that any multi-connected topology in which space was finite in at least one space direction had the effect of lowering the power spectrum at large wavelengths. Weeks et al. (2004) reexamined the question and showed that indeed, some finite multiconnected topologies do lower the large-scale fluctuations whereas others may elevate them. In fact, the long wavelengths modes tend to be relatively lowered only in a special family of closed multiconnected spaces called "well-proportioned". Generally, among spaces whose characteristic lengths are comparable with the radius of the last scattering surface $R_{lss}$ (a necessary condition for the topology to have an observable influence on the power spectrum), spaces with all dimensions of similar magnitude lower the quadrupole more heavily than the rest of the power spectrum. As soon as one of the characteristic lengths becomes significantly smaller or greater than the other two, the quadrupole is boosted in a way not compatible with WMAP data. In the case of flat tori, a cubic torus lowers the quadrupole whereas an oblate or a prolate torus increase the quadrupole; for spherical spaces, polyhedric spaces suppress the quadrupole whereas high order lens spaces (strongly anisotropic) boost the quadrupole. Thus, well-proportioned spaces match the WMAP data much better than the infinite flat space model.

## THE DODECAHEDRAL UNIVERSE

Among the family of well-proportioned spaces, the best fit to the observed power spectrum is the *Poincaré Dodecahedral Space*, hereafter PDS (Luminet et al., 2003). Recall that this space is positively curved, and is a multiconnected variant of the simply-connected hypersphere $S^3$, with a volume 120 times smaller for the same curvature radius. The associated power spectrum, namely the repartition of fluctuations as a function of their wavelengths corresponding to PDS, strongly depends on the value of the mass-energy density parameter. Luminet et al. (2003) computed the CMB multipoles for $l = 2, 3, 4$ and fitted the overall normalization factor to match the WMAP data at $l = 4$, and then examined their prediction for the quadrupole and the octopole as a function of $\Omega_0$. There is a small interval of values within which the spectral fit is excellent, and in agreement with the value of the total density parameter deduced from WMAP data (1.02 ± 0.02). The best fit is obtained for $\Omega_0 = 1.016$. Since then, the properties of PDS have been investigated in more details by various authors. Lachièze-Rey (2004) found an analytical expression of the eigenmodes of PDS, whereas Aurich et al. (2005) and Gundermann (2005) computed numerically the power spectrum up to the $l = 15$ mode and improved the fit with WMAP data. The result is quite remarkable because the Poincaré space has no degree of freedom. By contrast, a 3-dimensional torus, constructed by gluing together the opposite faces of a cube and which constitutes a possible topology for a finite Euclidean space, may be deformed into any parallelepiped : therefore its geometrical construction depends on 6 degrees of freedom.

The values of the matter density $\Omega_m$, of the dark energy density $\Omega_\Lambda$ and of the expansion rate $H_0$ fix the radius of the last scattering surface $R_{lss}$ as well as the curvature radius of space $R_c$, thus dictate the possibility to detect the topology or not. For $\Omega_m = 0.28$, $\Omega_0 = 1.016$ and $H_0 = 62$ *km/s/Mpc*, $R_{lss} = 53$ *Gpc* and $R_c = 2.63$ $R_{lss}$. It is to be noticed that the curvature radius $R_c$ is the same for the simply-connected universal covering space $S^3$ and for the multiconnected PDS. Incidently, the numbers above show that, contrary to a current opinion, a cosmological model with $\Omega_0 \sim 1.02$ is far from being "flat" (i.e. with $R_c = \infty$) ! For the same curvature radius than the simply-connected hypersphere $S^3$, the smallest dimension of the fundamental dodecahedron is only 43 Gpc, and its volume about 80 % the volume of the observable universe (namely the volume of the last scattering surface). This implies that some points of the last scattering surface will have several copies. Such a lens effect is purely attributable to topology and can be precisely calculated in the framework of the PDS model. It provides a definite signature of PDS topology, whereas the shape of the power spectrum gives only a hint for a small, well-proportioned universe model.

To be confirmed, the PDS model (soemtimes popularized as the "soccerball universe model") must satisfy two experimental tests :

1) New data from the future European satellite "Planck Surveyor" (scheduled 2007) could be able to determine the value of the energy density parameter with a precision of 1 %. A value lower than 1.009 would discard the Poincaré space as a model for cosmic space, in the sense that the size of the corresponding dodecahedron would become greater than the observable universe and would not leave any observable imprint on the CMB, whereas a value greater than 1.01 would strengthen its cosmological pertinence.

2) If space has a non trivial topology, there must be particular correlations in the CMB, namely pairs of "matched circles" along which temperature fluctuations should be the same (Cornish et al., 1998). The PDS model predicts 6 pairs of antipodal circles with an angular radius comprised between 5° and 55° (sensitively depending on the cosmological parameters).

Such circles have been searched in WMAP data by several teams, using various statistical indicators and massive computer calculations. First, Cornish et al. (2004) claimed to have found no matched circles on angular sizes greater than 25°, and thus rejected the PDS hypothesis. Next, Roukema et al. (2004) performed the same analysis for smaller circles, and found six pairs of matched circles distributed in a dodecahedral pattern, each circle on an angular size about 11°. This implies *$\Omega_0 = 1.010 \pm 0.001$ for $\Omega_m = 0.28 \pm 0.02$*, values which are perfectly consistent with the PDS model. Finally, Aurich et al. (2006a) performed a very careful search for matched circles and found that the putative topological signal in the WMAP data was considerably degraded by various effects, so that the dodecahedral space model could be neither confirmed nor rejected... This shows in passing how delicate the statistical analysis of observational data is, since different analyses of the same data can lead to radically opposed conclusions!

The controversy still went up a tone when Key et al. (2006) claimed that their negative analysis was not disputable, and that accordingly, not only the dodecahedral hypothesis was excluded, but also any multiply-connected topology on a scale smaller than the horizon radius. Since such an argument of authority, a fair portion of the academic community believes the WMAP data has ruled out multiply-connected models. However, at least the second part of the claim is wrong. The reason is that they searched only for antipodal or nearly-antipodal matched circles. But Riazuelo et al. (2004b) have shown that for generic multiply-connected topologies (including the well-proportioned ones, which are good candidates for explaining the WMAP power spectrum), the matched circles are generally not antipodal; moreover, the positions of the matched circles in the sky depend on the observer's position in the fundamental polyhedron. The corresponding larger number of degrees of freedom for the circles search in the WMAP data generates a dramatic increase of the computer time, up to values which are out-of-reach of the present facilities. It follows that the debate about the pertinence of PDS as the best fit to reproduce CMB observations is fully open.

The new release of WMAP data (Spergel et al., 2006), integrating two additional years of observation with reduced uncertainty, strengthened the evidence for an abnormally low quadrupole and other features which do not match with the infinite flat space model (this explains the unexpected delay in the delivery of this second release, originally announced for February 2004). Besides the quadrupole suppression, an anomalous alignment between the quadrupole and the octopole was put in evidence along a so-called « axis of evil » (Land and Magueijo, 2005). Thus the question arose to know whether, since non-trivial spatial topology can explain the weakness of the low-*l* modes, might it also explain the quadrupole-octopole alignment? Until then no multiply-connected space model, either flat (Cresswell et al., 2006) or spherical (Aurich et al., 2006b ; Weeks and Gundermann, 2006) was proved to exhibit the alignment observed in the CMB sky. This is not a strong argument against such models, since the « axis of evil » is generally interpreted as due to local effects and foreground contaminations (Prunet et al., 2005).

As a provisory conclusion, since some power spectrum anomalies are one of the possible signatures of a finite and multiply-connected universe, there is sill a continued interest in the Poincaré dodecahedral space and related finite universe models. And even if the particular dodecahedral space is eventually ruled out by future experiments, all of the other models of well-proportioned spaces will not be eliminated as such. In addition, numerical simulations show that, even if the size of a multiply-connected space is larger than that of the observable universe, we could all the same discover an imprint in the fossil radiation, even while no pair of circles, much less ghost galaxy images, would remain. The topology of the universe could therefore provide information on what happens outside of the cosmological horizon! But this is search for the next decade…

Maybe the most fundamental issue remains to link the present-day topology of space to a quantum origin, since classical general relativity does not allow for topological changes during the course of cosmic evolution. Theories of quantum gravity could allow to address the problem of a quantum origin of space topology. For instance, in the approach of quantum cosmology, some simplified solutions of Wheeler-de Witt equations show that the sum over all topologies involved in the calculation of the wavefunction of the universe is dominated by spaces with small volumes and multiconnected topologies (Carlip, 1993 ; e Costa & Fagundes, 2001). In the approach of brane worlds (see Brax 2003 for a review), the extra-dimensions are often assumed to form a compact Calabi-Yau manifold ; in such a case, it would be strange that only the ordinary dimensions of our 3-brane would not be compact like the extra ones. These are only heuristic indications on the way unified theories of gravity and quantum mechanics could "favor" multiconnected spaces. Whatsoever the fact that some particular multiconnected space models, such as PDS, may be refuted by future astronomical data, the question of cosmic topology will stay as a major question about the ultimate structure of our universe.